\documentclass[aps,prb,showpacs,showkeys]{revtex4}

\begin{document}
\title{A TOPOLOGICAL VIEW ON \\ BARYON NUMBER CONSERVATION}
\author{ERNESTO A. MATUTE}
\affiliation{Departamento de F\'{\i}sica \\ Universidad de
Santiago de Chile \\ Casilla 307 -- Correo 2, Santiago, Chile \\
ematute@lauca.usach.cl}

\begin{abstract}
We argue that the charge fractionalization in quarks has a hidden
topological character related to a broken ${\cal Z}_{2}$ symmetry
between integer-charged bare quarks and leptons.  The mechanism is
a tunneling process occurring in time between standard field
configurations of a pure gauge form with different topological
winding numbers associated with integer-charged bare quarks in the
far past and future.  This transition, which nonperturbatively
normalizes local bare charges with a universal accumulated value,
corresponds to a specific topologically nontrivial configuration
of the weak gauge fields in Euclidean spacetime. The outcome is an
effective topological charge equal to the ratio between baryon
number and the number of fermion generations associated with
baryonic matter.  The observed conservation of baryon number is
then related to the conservation of this bookkeeping charge on
quarks.  Baryon number violation may only arise through
topological effects as in decays induced by electroweak
instantons.  However, stability of a free proton is expected.
\end{abstract}


\maketitle

\renewcommand{\thesection}{\arabic{section}}
\renewcommand{\thetable}{\arabic{table}}

\section{Introduction}

Present experimental evidence indicates that baryons can be
produced only in particle-antiparticle pairs.\cite{PDG}  In
general, these particles are unstable and go through decay chains
that end with the lightest baryon, the proton.  A baryon number
B~=~1 is defined for a baryon and B~=~$-$1 for an
antibaryon.\cite{Lee} For all other particles the baryon number is
B~=~0.  A system of many particles has a baryon number given by
the corresponding algebraic sum.  The selection rule which
describes the observed baryon number conservation states that in
any particle reaction the baryon number is the same before and
after the interaction. Similar statements also apply to the
conservation of the lepton number~L. A necessary condition for the
baryon and lepton number conservation is proton stability.

In the quark model\cite{Close} all baryons are three-quark states
and each quark therefore has a baryon number B~=~1/3.  The
properties of quarks are correlated with the characteristics of
observed particle states.  So they occur in several diversities or
flavors distinguished by the assignment of additive quantum
numbers as in Table~\ref{Table1}.  The most unusual property of
quarks is that they have fractional charges. The baryon number
(B), flavor ($I_{z}$, $S$, $C$, $B^{*}$, or $T$), and electric
charge ($Q$) of quarks (and observed particles) are related
according to the generalized Gell-Mann--Nishijima
formula\cite{Perkins}
\begin{equation}
Q = I_{z} + \frac{1}{2} (\mbox{B} + S + C + B^{*} + T) .
\label{Nishi}
\end{equation}
The observed conservation of baryon number is then associated with
conservation of the number of quarks.

%-------------------------------------------------------------------
% TABLE 1
%-------------------------------------------------------------------
\begin{table}[t]
  \caption{\label{Table1} Quark additive quantum numbers.}
  \begin{ruledtabular}
  \begin{tabular}{lrrrrrr}
  Quark & u & d & s & c & b & t \\  \hline
  B (baryon number) & $\displaystyle \frac{1}{3}$ & $\displaystyle
  \frac{1}{3}$ & $\displaystyle \frac{1}{3}$ & $\displaystyle
  \frac{1}{3}$ & $\displaystyle \frac{1}{3}$ & $\displaystyle
  \frac{1}{3}$ \\
  & & & & & & \\
  $Q$ (electric charge) & $\displaystyle \frac{2}{3}$ & $\displaystyle
  - \frac{1}{3}$ & $\displaystyle - \frac{1}{3}$ & $\displaystyle
  \frac{2}{3}$ & $\displaystyle - \frac{1}{3}$ & $\displaystyle
  \frac{2}{3}$ \\
  & & & & & & \\
  $I_{z}$ (isospin $z$-component) & $\displaystyle \frac{1}{2}$ &
  $\displaystyle - \frac{1}{2}$ & 0 &
  0 & 0 & 0 \\
  & & & & & & \\
  $S$ (strangeness) & 0 & 0 & $-$1 & 0 & 0 & 0 \\
  & & & & & & \\
  $C$ (charm) & 0 & 0 & 0 & 1 & 0 & 0 \\
  & & & & & & \\
  $B^{*}$ (bottomness) & 0 & 0 & 0 & 0 & $-$1 & 0 \\
  & & & & & & \\
  $T$ (topness) & 0 & 0 & 0 & 0 & 0 & 1 \\
  \end{tabular}
  \end{ruledtabular}
\end{table}
%-------------------------------------------------------------------

During the past few decades, however, the baryon number and lepton
number conservation law has been questioned by the so-called grand
unified theories (GUTs) and supersymmetric grand unified theories
(SGUTs),\cite{GUTs} and some versions of string
theory.\cite{Strings} The reason essentially comes from gauge
theories which connect a charge conservation law with local gauge
invariance and the existence of appropriate massless fields.  No
known fields are related to baryon and lepton number conservation,
which are rather regarded as originating from an accidental global
Abelian gauge symmetry and therefore susceptible to be violated.
The main aim of GUTs is to unify all known interactions making
quarks and leptons in each generation share representations of the
unifying gauge group and consequently allowing direct transitions
between quarks and leptons.  Proton decay is one of the most
striking predictions of GUTs and their supersymmetric extensions,
having decay modes such as $p \rightarrow e^{+}+\pi^{o}$ or
$\bar{\nu}_{\mu}+K^{+}$ with $\Delta$B~=~$\Delta$L~=~$-$1.

The exact conservation law of baryon and lepton number has been
dismissed anyway as a consequence of electroweak instanton
effects.\cite{tHooft}  However, such violation effects of
topological character are unobservably small.  For example, while
the decay $p+n+n \rightarrow
e^{+}+\bar{\nu}_{\mu}+\bar{\nu}_{\tau}$ with
$\Delta$B~=~$\Delta$L~=~$-$3 is allowed, its probability is
extremely minuscule.

On the other hand, in the most popular extension of the standard model,
the minimal supersymmetric standard model (MSSM),\cite{MSSM} the
separate B and L conservation is usually enforced to maintain the
conservation of $R$-parity defined as $R=(-1)^{3(B-L)+2S}$ for a particle
of spin $S$.  If this parity is conserved, the number of supersymmetric
particles is then conserved.  On the contrary, non-minimal models of
low-energy supersymmetry with new interactions terms violating either
B or L and so $R$-parity conservation, but avoiding proton decay, have
also been discussed in the literature.\cite{foot1}

It is important therefore to have new strong theoretical arguments in
order to choose between these two opposite points of view for the
conservation of baryon and/or lepton number and the proton stability
question.  In this paper we argue that at the level of perturbation
theory the baryon number is conserved in any interaction coupling of
local fields due to a bookkeeping charge defined in terms of a conserved
effective topological charge on quarks.  We also show that there is
consistency with the nonperturbative topological effects that induce
baryon number violations. Our arguments are based on the so-called
Chern--Simons-fermion model of quarks\cite{Matute} recently proposed in
the context of the $\mbox{SU(3)}_{c} \mbox{xSU(2)}_{L} \mbox{xU(1)}_{Y}$
standard model of strong and electroweak interactions.\cite{Wein}  There
are two strong theoretical reasons for supposing that this new model
provides the appropriate framework for understanding the physics being
right beyond the standard model.  These are: (a)~it explains why quarks
have 1/3 electric charge relative to leptons, connecting this one-third
with the number of colors; (b)~it explains why quarks and leptons are so
similar in their weak-interaction properties, both occurring in
$\mbox{SU(2)}_{L}$ weak isospin doublets.  The noticeable aspect of the
approach is that these explanations are given within the scenario of the
standard model itself, without need to postulate new interactions as in
GUTs and preonic composite models.\cite{Harari}  Feature (a) is a
consequence of the nontrivial topological properties of weak gauge field
configurations which can nonperturbatively induce fractional charges
having a topological character; the topological charge or Pontryagin
index of non-Abelian fields is determined by the winding number
characterizing the fields at time infinity.\cite{Huang}  Thus, a quark
fractional \emph{bare} charge is seen as a result of the mixture between
an integer bare charge associated with a local field (which can be
derived from Noether's theorem) and a fractional topological charge
(which cannot be derived from Noether's theorem). It is assumed that only
integer bare charges are entirely associated with local fields, so that
such a topological charge structure does not exist in bare leptons.  This
topological viewpoint for fractional charges has also been advocated by
Wilczek\cite{Wilczek} on general grounds.  Topological charge
conservation therefore leads to an automatic conservation of baryon
number at the local field level.  Moreover, we find that in some sense
one is proportional to the other in baryonic matter.  Feature (b) is a
consequence of an electroweak ${\cal Z}_{2}$ symmetry between
integer-charged bare quarks and leptons.

In Sec.~2 we begin by considering the charge constraints which
indicate such a deeper discrete ${\cal Z}_{2}$ symmetry between
quarks and leptons.  This symmetry is then defined at a classical
bare level between integer-charged bare quarks and leptons.  In
Sec.~3 we discuss the counterterms needed to cancel the gauge
anomalies generated by the integer bare charges of quarks.  We
connect their nontrivial topological properties with the
fractional charge required to normalize nonperturbatively to quark
charge values.  The resulting charges are the standard bare quark
charges to begin with in the usual quantum field theory procedure.
In Sec.~4 a bookkeeping charge related to baryon number is defined
in terms of an effective topological charge on quarks. In Sec.~5
some concluding remarks are given.

\section{Integer-Charged Bare Quarks and the Hidden Quark-Lepton
Symmetry}

The starting point of our approach is the fact that the standard weak
hypercharge $Y$ of quarks and leptons are related as follows:
\begin{eqnarray}
& & Y(u_{L}) = Y(\nu_{eL}) + \frac{4}{3} ,  \hspace{1cm}
Y(u_{R}) = Y(\nu_{eR}) + \frac{4}{3} , \nonumber \\
& & \label{hyper} \\
& & Y(d_{L}) = Y(e_{L}) + \frac{4}{3} ,  \hspace{1cm} Y(d_{R}) =
Y(e_{R}) + \frac{4}{3} , \nonumber
\end{eqnarray}
and similarly for the other two generations, where the subscripts $L$ and
$R$ denote left- and right-handed components, respectively, and the
$\nu_{R}$'s are introduced because of the recent experimental signatures
for nonzero neutrino masses.\cite{PDG}  Also, since the hypercharge and
electric charge operators are chosen to be related according to the form
\begin{equation}
Q = T_{3} + \frac{1}{2} Y ,
\label{Q}
\end{equation}
where $T_{i}$ are the weak isospin generators, the electric charge of
quarks and leptons become connected by
\begin{equation}
Q(u) = Q(\nu_{e}) + \frac{2}{3} , \hspace{1cm}
Q(d) = Q(e) + \frac{2}{3} ,
\label{Qcharges}
\end{equation}
and similarly for the other generations.  The standard assignments of
weak isospin, hypercharge and electric charge are listed in
Table~\ref{Table2}.  Now it should be noted that the lepton hypercharges
are integer while the fractional shift for quarks, as seen from
Eq.~(\ref{hyper}), is 4/3 which depends neither on flavor nor on handness
and is conserved by electroweak interactions, i.e. one has the remarkable
result that the fractional hypercharge of a quark relies just on a
nonperturbative universal $4/3$ value that may have a topological
character:
%-------------------------------------------------------------------
% TABLE 2
%-------------------------------------------------------------------
\begin{table}[b]
  \caption{\label{Table2} Weak isospin, hypercharge and electric charge of
  the first generation of quarks and leptons.}
  \begin{ruledtabular}
  \begin{tabular}{lrrrrrrrr}
  Fermion & u$_{L}$ & d$_{L}$ & u$_{R}$ & d$_{R}$
  & $\nu_{eL}$ & e$_{L}$ & $\nu_{eR}$ & e$_{R}$    \\  \hline
  $T$ & $\displaystyle \frac{1}{2}$ & $\displaystyle \frac{1}{2}$ &
  0 & 0 & $\displaystyle \frac{1}{2}$ & $\displaystyle \frac{1}{2}$ &
  0 & 0 \\
  & & & & & & & & \\
  $T_{3}$ & $\displaystyle \frac{1}{2}$ & $\displaystyle - \frac{1}{2}$ &
  0 & 0 & $\displaystyle \frac{1}{2}$ & $\displaystyle - \frac{1}{2}$ &
  0 & 0 \\
  & & & & & & & & \\
  $Y$ & $\displaystyle \frac{1}{3}$ & $\displaystyle \frac{1}{3}$ &
  $\displaystyle \frac{4}{3}$ & $\displaystyle - \frac{2}{3}$
  & $-$1 & $-$1 & 0 & $-$2 \\
  & & & & & & & & \\
  $Q$ & $\displaystyle \frac{2}{3}$ & $\displaystyle - \frac{1}{3}$ &
  $\displaystyle \frac{2}{3}$ & $\displaystyle - \frac{1}{3}$
  & 0 & $-$1 & 0 & $-$1 \\
  \end{tabular}
  \end{ruledtabular}
\end{table}
%-------------------------------------------------------------------
\begin{equation}
g' Y(q) = g' (m + \frac{4}{3}) , \label{Yq}
\end{equation}
where $m$ is an integer having values $0, -1, -2$ depending on
quark flavor and handness, and $g'$ is the U(1)$_{Y}$ coupling.
Hence, to understand these patterns, it is natural to take the
view that Eqs.~(\ref{hyper}) and (\ref{Yq}) are a manifestation of
a hidden electroweak ${\cal Z}_{2}$ symmetry between quarks and
leptons as well as an indication for the fractional hypercharge of
an underlying field configuration in bare quarks that may have
topological attributes.  This presumption was discussed in
Ref.~\onlinecite{Matute} within the framework of the
Chern--Simons-fermion model of quarks.  In general, the
topological character of fractional charges has also been
emphasized in Ref.~\onlinecite{Wilczek}.  We remark that an
alternative dynamical description would be to associate the
fractional 4/3 hypercharge with a fundamental preon and construct
an appropriate composite model with new binding
forces.\cite{Harari}  However, it is shown that the quark
fractional hypercharge in Eq.~(\ref{Yq}) can be accounted for only
using the topological properties of certain standard gauge field
configurations, so that a preon model is not necessary.

According to the mentioned topological approach to fractional
hypercharge, bare quarks may be viewed as fermions with integer
bare hypercharges, referred to as prequarks, which are
nonperturbatively normalized with a universal fractional piece
generated by a topologically nontrivial Euclidean configuration of
gauge fields.  In other words, a self-consistent method that
adjusts prequark hypercharges with a universal Chern--Simons
contribution is pursued at the classical field theory level.  Bare
leptons, which have integer hypercharges, are supposed to receive
no such a topological charge contribution at all. Thus, if we
denote a prequark with a hat over the symbol representing the
appropriate quark and the Euclidean gauge field configuration
with~$X$, bare quarks may conveniently be looked upon as ``made''
of the following mixtures:
\begin{eqnarray}
& & u = \{\hat{u} X\} , \hspace{1cm} c = \{\hat{c} X\} , \hspace{1cm}
t = \{\hat{t} X\} , \nonumber \\ & & \label{composite} \\
& & d = \{\hat{d} X\} , \hspace{1cm} s = \{\hat{s} X\} ,
\hspace{1cm} b = \{\hat{b} X\} . \nonumber
\end{eqnarray}
Of course, these structures are not authentic in the sense that gauge
fields really live in Minkowski spacetime.  Also, no new binding forces
are involved.  We argue below that in Minkowski spacetime the
$X$-configuration is a tunneling process occurring in time by which
prequarks get fractional charges.  The situation is similar to the one
found in instanton physics:\cite{tHooft} on the one hand, an instanton
is a localized pseudoparticle in Euclidean spacetime, on the other hand,
it is a tunneling process in Minkowski spacetime.

Each prequark has the spin, isospin, color charge, flavor, and weak
isospin of the corresponding quark; its bare weak hypercharge is the same
as its lepton partner (see Eq.~(\ref{hyper})) and its bare electric
charge is defined by Eq.~(\ref{Q}), which in turn fixes its bare baryon
number through Eq.~(\ref{Nishi}).  Prequark charges are bare charges as
prequarks are nonperturbatively dressed into bare quarks at the
classical field theory level.  Tables~\ref{Table3} and \ref{Table4} list
the prequark quantum numbers.

%-------------------------------------------------------------------
% TABLE 3
%-------------------------------------------------------------------
\begin{table}[h]
  \caption{\label{Table3} Prequark additive quantum numbers.}
  \begin{ruledtabular}
  \begin{tabular}{lrrrrrr}
  Prequark & $\hat{u}$ & $\hat{d}$ & $\hat{s}$ & $\hat{c}$ &
  $\hat{b}$ & $\hat{t}$  \\  \hline
  B (baryon number) & $-$1 & $-$1 & $-$1 & $-$1 & $-$1 & $-$1 \\
  & & & & & & \\
  $Q$ (electric charge) & 0 & $-$1 & $-$1 & 0 & $-$1 & 0 \\
  & & & & & & \\
  $I_{z}$ (isospin $z$-component) & $\displaystyle \frac{1}{2}$ &
  $\displaystyle - \frac{1}{2}$ & 0 &
  0 & 0 & 0 \\
  & & & & & & \\
  $S$ (strangeness) & 0 & 0 & $-$1 & 0 & 0 & 0 \\
  & & & & & & \\
  $C$ (charm) & 0 & 0 & 0 & 1 & 0 & 0 \\
  & & & & & & \\
  $B^{*}$ (bottomness) & 0 & 0 & 0 & 0 & $-$1 & 0 \\
  & & & & & & \\
  $T$ (topness) & 0 & 0 & 0 & 0 & 0 & 1 \\
  \end{tabular}
  \end{ruledtabular}
\end{table}
%-------------------------------------------------------------------

%-------------------------------------------------------------------
% TABLE 4
%-------------------------------------------------------------------
\begin{table}[h]
  \caption{\label{Table4} Weak isospin, hypercharge and electric charge of
  the first generation of prequarks.}
  \begin{ruledtabular}
  \begin{tabular}{lrrrr}
  Prequark & $\hat{u}_{L}$ & $\hat{d}_{L}$ & $\hat{u}_{R}$ &
  $\hat{d}_{R}$  \\  \hline
  $T$ & $\displaystyle \frac{1}{2}$ & $\displaystyle \frac{1}{2}$ &
  0 & 0 \\
  & & & & \\
  $T_{3}$ & $\displaystyle \frac{1}{2}$ & $\displaystyle - \frac{1}{2}$
  & 0 & 0 \\
  & & & & \\
  $Y$ & $-$1 & $-$1 & 0 & $-$2 \\
  & & & & \\
  $Q$ & 0 & $-$1 & 0 & $-$1 \\
  \end{tabular}
  \end{ruledtabular}
\end{table}
%-------------------------------------------------------------------

Now regarding the bare quantum numbers associated with the
$X$-configuration of gauge fields, from Eqs.~(\ref{hyper}), (\ref{Q})
and (\ref{Nishi}) we get
\begin{equation}
Y(X) = \frac{4}{3} , \hspace{1cm}
Q(X) = \frac{1}{2} Y(X) = \frac{2}{3} , \hspace{1cm}
\mbox{B}(X) = 2 Q(X) = \frac{4}{3} .
\label{Xcharges}
\end{equation}
When these fractional charges are added to the integer prequark ones,
the fractional quark bare charges are obtained (see
Eq.~(\ref{composite}) and Tables~\ref{Table1}--\ref{Table4}).
In particular, the field configuration interpolates between the
prequark and quark baryon numbers. It clearly distinguishes quarks from
leptons and therefore it does not allow baryon number violating
fundamental couplings; i.e. the conservation law of baryon number in
classical field theory is an automatic consequence of the topologically
nontrivial gauge field configuration underlying bare quarks.

On the other hand, the underlying topological structure of quark
fractional bare charges uncovers an electroweak ${\cal Z}_{2}$ symmetry
between bare prequarks and leptons, which we refer to as presymmetry.
This symmetry means invariance of the classical electroweak Lagrangian of
prequarks and leptons under the transformation
\begin{equation}
 \hat{u}^{i}_{L} \leftrightarrow \nu_{eL} , \hspace{1cm}
\hat{u}^{i}_{R} \leftrightarrow \nu_{eR} , \hspace{1cm}
\hat{d}^{i}_{L} \leftrightarrow e_{L} , \hspace{1cm}
\hat{d}^{i}_{R} \leftrightarrow e_{R} ,
\end{equation}
and similarly for the other generations, where $i$ denotes the
color degree of freedom (see Tables~\ref{Table2} and
\ref{Table4}).  We observe that prequarks and leptons have the
same B~$-$~L~=~$-$1; i.e. B~$-$~L is the right fermion number to
be considered under presymmetry.\cite{foot2} Besides, it is
important to realize that presymmetry is not an \emph{ad hoc}
symmetry; as Eq.~(\ref{hyper}) shows, it underlies the
relationships between quarks and leptons in the electroweak sector
of the standard model.

\section{Counterterms and Bare Charge Normalization}

We now explain how the field configuration $X$ gets effectively involved
in the local dynamics of flavors.  We also prove self-consistently that
quarks have fractional charges according to the quantum numbers given
above.  Although we follow Ref.~\onlinecite{Matute}, new insights into
the classical dynamics of the standard gauge fields are presented.

In Minkowski spacetime the current related to the $X$-configuration of
gauge fields is defined by the gauge-dependent form\cite{Matute}
\begin{eqnarray}
J^{\mu}_{X} &=& \frac{1}{4 N_{\hat{q}}} K^{\mu}
\sum_{\hat{q}_{L} , \ell_{L}} Y + \frac{1}{16 N_{\hat{q}}} L^{\mu}
\left( \sum_{\hat{q}_{L} , \ell_{L}} Y^{3} -
\sum_{\hat{q}_{R} , \ell_{R}} Y^{3} \right) \nonumber \\
& & \nonumber \\
&=& - \frac{1}{6} \, K^{\mu} + \frac{1}{8} \, L^{\mu} ,
\label{currX}
\\ & & \nonumber
\end{eqnarray}
where the sums are over all the fermion representations, $N_{\hat{q}}$=36
for three generations of prequarks, and
\begin{eqnarray}
K^{\mu} &=& \frac{g^{2}}{8 \pi^{2}} \epsilon^{\mu\nu\lambda\rho}
tr \left( W_{\nu} \partial_{\lambda} W_{\rho} - \frac{2}{3} i g W_{\nu}
W_{\lambda} W_{\rho} \right) , \nonumber \\
& & \label{CS} \\
L^{\mu} &=& \frac{{g'}^{2}}{12 \pi^{2}}
\epsilon^{\mu\nu\lambda\rho} A_{\nu} \partial_{\lambda} A_{\rho} ,
\nonumber
\end{eqnarray}
which are the Chern--Simons classes or topological currents connected
with the $\mbox{SU(2)}_{L}$ and $\mbox{U(1)}_{Y}$ gauge groups of the
standard model, respectively.  Their interactions are dictated by the
presymmetric Lagrangian
\begin{equation}
{\cal L} = g' N_{\hat{q}} \, J^{\mu}_{X} A_{\mu} ,
\label{Lagran}
\end{equation}
so that only the non-Abelian fields are topologically relevant, as
expected.  It is worth noting here that the $X$-configuration carries
hypercharge, but no color and flavor.  Besides, this gauge-dependent
local counterterm is added to the classical Lagrangian of massless
prequarks and leptons in order to remove the $\mbox{U(1)x[SU(2)]}^{2}$
and $\mbox{[U(1)]}^{3}$ gauge anomalies generated by the integer charge
of prequarks and so restore gauge invariance in a quantum field theory
scenario.  In fact, the U(1) current for prequarks and leptons in all
representations
\begin{equation}
\hat{J}^{\mu}_{Y} = \overline{\hat{q}}_{L} \gamma^{\mu} \frac{Y}{2}
\hat{q}_{L} + \overline{\hat{q}}_{R} \gamma^{\mu} \frac{Y}{2}
\hat{q}_{R} + \overline{\ell}_{L} \gamma^{\mu} \frac{Y}{2} \ell_{L} +
\overline{\ell}_{R} \gamma^{\mu} \frac{Y}{2} \ell_{R} ,
\end{equation}
is gauge invariant but it is not conserved.  Instead, due to the
mentioned triangle gauge anomalies,
\begin{equation}
\partial_{\mu} \hat{J}^{\mu}_{Y} = - N_{\hat{q}} \; \partial_{\mu}
J^{\mu}_{X} ,
\end{equation}
where $J^{\mu}_{X}$ is the current given in Eq.~(\ref{currX}).

The Lagrangian in Eq.~(\ref{Lagran}), however, allows us to define a new
bare current:
\begin{equation}
J^{\mu}_{Y} = \hat{J}^{\mu}_{Y} + N_{\hat{q}} \; J^{\mu}_{X} ,
\label{newJ}
\end{equation}
which is conserved but it is gauge dependent.  The corresponding
conserved charge in Minkowski spacetime
\begin{equation}
Q_{Y}(t) = \int d^{3}x \; J^{o}_{Y}
\label{Qt}
\end{equation}
is also gauge-dependent.  Nevertheless, it has a nontrivial
topological character which allows to solve all problems at once.
Indeed, presymmetry has available a degree of freedom which, as
shown below, can be chosen conveniently through Eq.~(\ref{Qt}) to
restore gauge invariance.  Of course, a fixed value breaks such a
symmetry.  The parameter we are referring to is the topological
charge or Pontryagin index.

Actually, charge in Eq. (\ref{Qt}) is not conserved because of the
topological charge associated with gauge fields.  To see how this is
possible, we calculate the change in $Q_{Y}$ between $t = - \infty$ and
$t = + \infty$.  We assume, as usual, that the region of spacetime where
the energy density is nonzero is bounded.\cite{Jackiw,Callan}  Therefore
this region can be surrounded by a three-dimensional surface on which the
field configuration becomes pure gauge, i.e.
\begin{equation}
W_{\mu} = - \frac{i}{g} (\partial_{\mu} U) U^{-1}  \label{puregauge}
\end{equation}
at the boundary, which can be obtained from $W_{\mu}=0$ by a
transformation $U$ that takes values in the corresponding gauge group.
It is convenient to use Eqs.~(\ref{puregauge}), (\ref{CS}),
(\ref{currX}), (\ref{newJ}) and (\ref{Qt}) just for the pure gauge
fields.  In this case we end up with
\begin{equation}
Q_{Y}(t) = \frac{N_{\hat{q}}}{6} \; n_{W}(t) ,
\end{equation}
where
\begin{equation}
n_{W}(t) = \frac{1}{24 \pi^2} \int d^{3}x \; \epsilon^{ijk}
tr (\partial_{i}U U^{-1} \partial_{j}U U^{-1} \partial_{k}U U^{-1})
\label{winding}
\end{equation}
is the winding number of the non-Abelian gauge transformation.  This
number is integer-valued if we consider a fixed time $t$ and assume that
$U(t,{\bf x})$ equals a direction independent constant at spatial
infinity, e.g. $U \rightarrow 1$ for $|{\bf x}| \rightarrow \infty$.
The usual argument to see this property is based on the
observation that this $U$ can be viewed as a map from the
three-dimensional space with all points at infinity regarded as the same
onto the three-dimensional sphere of parameters $S^{3}$ of the
$\mbox{SU(2)}_{L}$ group manifold.  But three-space with all points at
infinity being in fact one point is topologically equivalent to a sphere
$S^{3}$ in Minkowski space.  Therefore $U$ determines a map $S^{3}
\rightarrow S^{3}$.  These maps can be characterized by an integer
topological index which labels the homotopy class of the map.  This
integer is analytically given by Eq.~(\ref{winding}).  For the Abelian
case, $n_{W}=0$ since $A_{\mu}=\partial_{\mu}\chi$, $U=e^{ig'\chi}$,
$\partial_{k}U U^{-1} = ig'\partial_{k}\chi$, and $\epsilon^{ijk}
\partial_{i}\chi \partial_{j}\chi \partial_{k}\chi =0$.

Thus, for the field configurations joined to prequarks which at the
initial $t = - \infty$ and the final $t = + \infty$ are supposed to be
of the above pure gauge form with different winding numbers, the change
in charge becomes
\begin{eqnarray}
\Delta Q_{Y} &=& Q_{Y}(t=+\infty) - Q_{Y}(t=-\infty) \nonumber \\
& & \nonumber \\
&=& \frac{N_{\hat{q}}}{6} \left[ n_{W}(t=+\infty) - n_{W}(t=-\infty)
\right] .
\label{deltaQ1}
\end{eqnarray}
The difference between the integral winding numbers of the pure gauge
configurations characterizing the gauge fields at the far past and future
can be rewritten as the topological charge or Pontryagin index defined in
Minkowski spacetime by
\begin{equation}
Q_{T} = \int d^{4}x \; \partial_{\mu} K^{\mu} = \frac{g^{2}}{16 \pi^{2}}
\int d^{4}x \; tr (W_{\mu\nu} \tilde{W}^{\mu\nu}) ,
\label{QT1}
\end{equation}
where it is assumed that $K^{i}$ decreases rapidly enough at spatial
infinity and
\begin{eqnarray}
& & W_{\mu\nu} = \tau^{a} W^{a}_{\mu\nu} , \nonumber \\
& & \nonumber \\
& & \tilde{W}_{\mu\nu} = \frac{1}{2} \epsilon_{\mu\nu\lambda\rho}
\tau^{a} W^{a \lambda\rho} , \\
& & \nonumber \\
& & W^{a}_{\mu\nu} = \partial_{\mu} W^{a}_{\nu} -
\partial_{\nu} W^{a}_{\mu} + g \epsilon^{abc} W^{b}_{\mu} W^{c}_{\nu} .
\nonumber
\end{eqnarray}
It is gauge invariant, conserved and, for arbitrary fields, can take any
real value.  But, as shown above, for a pure gauge configuration it is
integer-valued, so that using Eq.~(\ref{deltaQ1}) we are led to
\begin{equation}
Q_{T} = n = n_{W}(t=+\infty) - n_{W}(t=-\infty) ,        \label{QT2}
\end{equation}
and then
\begin{equation}
\Delta Q_{Y} = N_{\hat{q}} \; \frac{n}{6} .              \label{deltaQ2}
\end{equation}
These equations have a special significance when the integral in
Eq.~(\ref{QT1}) is analytically continued to Euclidean spacetime.  The
integral topological charge remains the same but it can now be associated
with an Euclidean field configuration that at this point we identify with
the $X$ in Eq.~(\ref{composite}).  In this case, the meaning of
Eq.~(\ref{QT2}) is that the Euclidean $X$-configuration interpolates in
imaginary time between the real time pure-gauge configurations in the far
past and future which are topologically inequivalent as they have
different winding numbers.  The condition is like the one established for
instantons.  Thus the continuous interpolation is to be considered as a
tunneling process, so that a barrier must separate the initial and final
gauge field configurations.  We further note that if such an analytical
continuation to Euclidean spacetime is ignored, nonzero topological
charge then implies nonvanishing energy density at intermediate real time
and so no conservation of energy.  At the end of the process, there is an
accumulated hypercharge on prequarks given in Eq. (\ref{deltaQ2}).  So
the charge of a prequark does not only depend on fields at a single
instant of time, but also on flow of current at infinity.  Specifically,
Eq.~(\ref{deltaQ2}) means that $N_{\hat{q}}$ prequarks have to change
their $Y/2$ in the same amount $n/6$; bare leptons have integer charge
with no topological contribution from the gauge field configuration.  For
each prequark it implies the hypercharge change
\begin{equation}
Y_{\hat{q}} \rightarrow Y_{\hat{q}}  + \frac{n}{3} .     \label{norma}
\end{equation}
Therefore the nontrivial topological properties of the $X$-configuration
give an extra contribution to prequark hypercharge, so that the integer
bare values we start with have to be shifted.  Accordingly, the gauge
anomalies have to be re-evaluated.  It is found that anomalies are
cancelled self-consistently for $n=4$, the number of prequark flavors per
generation, since now
\begin{equation}
\sum_{q_{L}, \ell_{L}} Y = 0 , \hspace{1cm}
\sum_{q_{L} , \ell_{L}} Y^{3} - \sum_{q_{R} , \ell_{R}} Y^{3} = 0 .
\label{sumsY}
\end{equation}

The above nonperturbative hypercharge normalization with
topological charge $n=4$ is consistent with Eq.~(\ref{Yq}) and it
means restoration of gauge invariance, breaking of the electroweak
presymmetry in the Abelian sector, dressing of prequarks into
fractional-charged bare quarks, and the replacement of the bare
presymmetric model by the standard model.  In fact, from
Eqs.~(\ref{sumsY}) and (\ref{currX}) we note that the
gauge-dependent topological current $J^{\mu}_{X}$ related to the
$X$-configuration is canceled.  However, the corresponding
conserved topological charge is gauge-independent and manifests
itself as a universal part of prequark hypercharges according to
Eq.~(\ref{Yq}). More precisely, if we consider the topological
current $N_{\hat{q}} J^{\mu}_{X}$ in Eq.~(\ref{newJ}) and its
associated $X$-configuration causing the hypercharge obtained in
Eq.~(\ref{deltaQ2}), we may define an effective prequark local
current $\hat{J}^{\mu}_{Y,\mbox{\footnotesize{eff}}}$ by
\begin{equation}
\hat{J}^{\mu}_{Y,\mbox{\footnotesize{eff}}} = \frac{2}{3} \left(
\overline{\hat{q}}_{L} \gamma^{\mu} \hat{q}_{L} +
\overline{\hat{q}}_{R} \gamma^{\mu} \hat{q}_{R} \right)
\end{equation}
to absorb their effects, namely, anomaly cancellation and induced
hypercharge $4/3$ on prequarks regardless of flavor and handness.
Thus Eq.~(\ref{newJ}) takes the gauge-independent form
\begin{equation}
J^{\mu}_{Y} = \overline{\hat{q}}_{L} \gamma^{\mu} \frac{Y+4/3}{2}
\hat{q}_{L} + \overline{\hat{q}}_{R} \gamma^{\mu} \frac{Y+4/3}{2}
\hat{q}_{R}
+ \overline{\ell}_{L} \gamma^{\mu} \frac{Y}{2} \ell_{L} +
\overline{\ell}_{R} \gamma^{\mu} \frac{Y}{2} \ell_{R} .
\end{equation}
At this point, prequarks with normalized local hypercharge, which
effectively includes the nonperturbative universal $4/3$ part, have to
be identified with standard bare quarks.  The replacement of prequarks by
quarks in the strong, weak and Yukawa sectors is straightforward as they
have the same color, flavor and weak isospin.

All of this is essentially done at the level of the classical Lagrangian
where the current $J^{\mu}_{X}$, which only mixes with prequark currents,
is used.  Again, the gauge configuration $X$ related to this topological
current in the far past and future is a pseudoparticle in Euclidean
spacetime and a tunneling process in Minkowski spacetime by which a
prequark hypercharge nonperturbatively changes from integer to
fractional values.  It is also interesting to note that as stated by the
model self-consistency is the reason for the ``magical'' cancellation of
gauge anomalies in the standard model.  Moreover, the factor 1/3 in
Eq.~(\ref{norma}) is due to the number of prequark colors (assuming same
number of prequark and lepton families) introduced in Eq.~(\ref{currX})
through $N_{\hat{q}}$, which predicts, as expected, that quarks carry
1/3-integral charge because they have three colors.

Our analysis shows that quarks instead of prequarks are the fermions to
start with in the quantum field theory treatment.  The novel news is that
bare quarks have fractional charges owing to a nonperturbative universal
contribution from an instanton-like classical gauge field configuration
with specific topological properties (i.e. $n=4$).  And underlying this
charge structure one has presymmetry as reflected in Eq.~(\ref{hyper});
an electroweak ${\cal Z}_{2}$ symmetry between bare quarks and leptons
which is broken by the vacuum configuration of gauge fields but it
accounts for the electroweak similarities between quarks and leptons.
In this sense, we recall that starting with bare fermions and exact
symmetries, and ending with normalized fermions and broken or hidden
symmetries is a conventional procedure in field theory which allows to
understand properties of matter and interactions.

\section{Bookkeeping Charge on Baryons}

Next we remark that an effective fractional topological charge equal to
$Q_{T}=n/N_{\hat{q}}=1/N_{c}N_{g}=1/9$, where $N_{c}$ and $N_{g}$ are the
numbers of colors and generations of prequarks, respectively, can be
associated with each $X$-configuration.  Within the classical
configuration of Eq.~(\ref{composite}), it is then natural to define an
effective topological charge for quarks, a bookkeeping charge which is
conserved and should be related to its fermion number.  As a general
rule we find that
\begin{equation}
Q_{T} = \frac{\mbox{B}}{N_{g}} .
\label{rule}
\end{equation}
Thus $\Delta$B = $N_{g} \Delta Q_{T}$ for any baryon-number violating
process, i.e. only topological effects may violate baryon number.

To see the consistency of the definition for the bookkeeping
charge in Eq.~(\ref{rule}), we consider the baryon plus lepton
number (B~+~L) violating processes induced nonperturbatively by
electroweak instanton effects.  According to
Ref.~\onlinecite{tHooft}, for three generations, one electroweak
instanton characterized by a topological charge $Q_{T}=1$ is
associated with quark and lepton number violations given by
\begin{eqnarray}
& & \Delta u + \Delta d = -3 , \hspace{1cm} \Delta e + \Delta
\nu_{e} = -1 , \nonumber \\ & & \nonumber \\ & & \Delta c + \Delta
s = -3 , \hspace{1cm} \Delta \mu + \Delta \nu_{\mu} = -1 ,
\label{Deltas} \\ & & \nonumber \\ & & \Delta t + \Delta b = -3 ,
\hspace{1cm} \Delta \tau + \Delta \nu_{\tau} = -1 . \nonumber
\end{eqnarray}
In total, the baryon and lepton numbers are violated in three units
($\Delta$B = $\Delta$L = $-$3), matching three baryons or nine quarks
($d$, $s$, and $b$ may mix through the Kobayashi--Maskawa matrix) and
three antileptons.  A decay such as
$p+n+n \rightarrow \mu^{+}+\bar{\nu}_{e}+\bar{\nu}_{\tau}$ is then
allowed.  Within the classical configuration scheme of
Eq.~(\ref{composite}), Eq.~(\ref{Deltas}) implies
\begin{eqnarray}
& & \Delta \hat{u} + \Delta \hat{d} = -3 , \hspace{1cm}  \Delta X = -3 ,
\nonumber \\ & & \nonumber \\
& & \Delta \hat{c} + \Delta \hat{s} = -3 , \hspace{1cm}  \Delta X = -3 ,
\\ & & \nonumber \\
& & \Delta \hat{t} + \Delta \hat{b} = -3 , \hspace{1cm}  \Delta X = -3 ,
\nonumber
\end{eqnarray}
i.e. one instanton induces a process in which nine $X$-configurations
vanish.  But the nine $X$'s make precisely the same topological charge
of the electroweak instanton.  In a sense, the definition in
Eq.~(\ref{rule}) therefore brings back topological charge conservation in
quantum flavor dynamics.  In Minkowski spacetime one electroweak
instanton corresponds to a process which has associated the topological
charge change $\Delta Q_{T} = 1$.  It induces a process with the
effective topological charge change $\Delta Q_{T} = -1$ associated with
the vanishing of nine quarks and baryon number violation $\Delta \mbox{B}
= -3$. We further note that consistency between the topological charge of
the instanton and the effective one assigned to quarks corroborates the
above value $n=4$ of the Pontryagin index fixed by gauge anomaly
constraints.

Finally, it can be seen from Eq.~(\ref{rule}) that a baryon number
violation  $\Delta$B~=~$-$1 in actual experiments means an
effective topological-charge change $\Delta Q_{T} = - 1 / N_{g}$.
In particular, proton decay would imply the presence of a
background gauge source with topological charge $1 /
N_{g}$.\cite{foot3} Stability of a free proton is expected anyway
because instanton-like events cannot change topological charge by
this amount.  For three generations, the effective topological
charge has a value $Q_{T}=1/9$ for quarks and $Q_{T}=1/3$ for
nucleons.

\section{Conclusion}

In this paper we have considered a discrete ${\cal Z}_{2}$ symmetry
between integer-charged bare quarks and leptons to understand the
quark--lepton similarities in the electroweak sector of the standard
model.  New insights into the classical dynamics of the weak gauge fields
which nonperturbatively give to bare quarks fractional charge relative
to bare leptons have been provided.  A self-consistent method to adjust
bare charge values with Chern--Simons topological contributions has been
pursued; it is assumed that only integer bare charges are entirely
associated with local fields.

We have also presented arguments to explain the observed conservation of
baryon number and predict stability of a free proton, based on the
conservation of a bookkeeping charge defined in terms of an effective
topological charge on quarks.  They contravene GUTs and their
supersymmetric extensions, where baryon number violation couplings are
introduced from the beginning, i.e. at the level of the classical
Lagrangian.  However, they support the MSSM and its non-minimal
extensions with baryon number conservation, which then appear as the
most promising approachs to low-energy supersymmetry; it is worth
remarking here that while bare quarks and their supersymmetric partners
have the same baryon number and underlying topological charge
configuration, bare leptons and their superpartners, which carry integer
charges, have zero baryon number and no such a topological charge
structure at all.  Of course, all of this essentially applies within the
classical field theory analysis, which ends with the known standard bare
fermions in a self-consistent way.  These are the ones to start with in
the usual quantum field theory study.

\begin{acknowledgements}
We would like to thank J. Gamboa and L. Vergara for valuable discussions.
This work was partially supported by the Departamento de Investigaciones
Cient\'{\i}ficas y Tecnol\'ogicas, Universidad de Santiago de Chile.
\end{acknowledgements}


\begin{thebibliography}{99}
\bibitem{foot2} The baryon number of prequarks may be set equal
to~1 if presymmetry between prequarks and antileptons
($\hat{u}^{i}~\leftrightarrow~e^{c}$,
$\hat{d}^{i}~\leftrightarrow~\nu^{c}$)
is assumed in the model, with B$-$L=1 for the exchanged
fermions.  Also, $Y(X)=-2/3$, $Q(X)=-1/3$ and $\mbox{B}(X)=-2/3$,
instead of the charge values given in Eq.~(\ref{Xcharges}).
\end{thebibliography}
\end{document}